\documentclass[aps,prapplied,reprint,superscriptaddress,showkeys,footinbib,floatfix]{revtex4-2}

\usepackage[T1]{fontenc}				
\usepackage[utf8]{inputenc}			
\usepackage[english]{babel}			
\usepackage[protrusion=false]{microtype}	

\usepackage{color} 						
\definecolor{red}{rgb}{1,0,0}				
\definecolor{blue}{rgb}{0,0,1}				
\definecolor{black}{rgb}{0,0,0}				

\usepackage{soul} 
\definecolor{hlyellow}{rgb}{0.95,0.95,0}
\definecolor{hlgreen}{rgb}{0,0.95,0}
\sethlcolor{hlyellow}
\soulregister \ref 7
\soulregister \cite 7
\soulregister \citet 7
\soulregister \footnote 8
\soulregister {\ } 7
\soulregister \figref 7
\soulregister \tabref 7
\soulregister \secref 7
\soulregister \refref 7
\soulregister \eqref 7

\usepackage{amsmath} 
\usepackage{amssymb,amsfonts,mathrsfs} 
\usepackage{array} 

\usepackage{graphicx} 

\usepackage{hyperref} 
\definecolor{dullmagenta}{rgb}{0.4,0,0.4} 
\definecolor{darkblue}{rgb}{0,0,0.4}
\definecolor{medblue}{rgb}{0,0,0.6}
\definecolor{lightblue}{rgb}{0,0,0.8}
\hypersetup{
	colorlinks=true, 		
	breaklinks=true,		
	linkcolor=lightblue,
	citecolor=lightblue,
	urlcolor=lightblue,
	filecolor=dullmagenta
}
\usepackage[all]{hypcap} 

\usepackage{fixgreek}
\usepackage{physmaths}
\usepackage{physunits}

\newcommand{\figpdfa}{}
\newcommand{\figref}[1]{Fig.\ \ref{#1}} 
\newcommand{\tabref}[1]{Tab.\ \ref{#1}} 
\newcommand{\secref}[1]{Sec.\ \ref{#1}} 
\newcommand{\refref}[1]{Ref.\ \cite{#1}} 
\newcommand{\eqnref}[1]{Eq.\ \eqref{#1}} 
\newcommand{\figletter}[1]{\textbf{(\mbox{#1})}}

\newcommand{\PHC}{\mbox{P.H.-C.}}

\newcommand{\hyphenohm}{\text{-}\Omega}				
\newcommand{\hyphenkohm}{\text{-}\text{k} \Omega}			
\newcommand{\hyphenmm}{\text{-}\text{mm}}		
\newcommand{\hyphenGHz}{\text{-}\text{GHz}}		
\newcommand{\hyphennm}{\text{-}\text{nm}}			

\newcommand{\appref}[1]{Appendix \ref{#1}} 
\newcommand{\SupRefSecFab}{\appref{sec:fabrication}} 
\newcommand{\SupRefSecModel}{\appref{sec:model}} 
\newcommand{\SupRefSecSetup}{\appref{sec:setup}} 
\newcommand{\SupRefSecAnalysis}{\appref{sec:analysis}} 

\addto\captionsenglish{%
  \renewcommand{\appendixname}%
    {APPENDIX}%
}
\begin{document}

\newcommand{\mytitle}
{On-chip microwave filters for high-impedance resonators with gate-defined \\quantum dots}
\title{\mytitle}

\author{Patrick \surname{Harvey-Collard}}
\email[Correspondence to: ]{P.Collard@USherbrooke.ca}
\affiliation{QuTech and Kavli Institute of Nanoscience, Delft University of Technology, 2628 CJ Delft, The Netherlands}

\author{Guoji \surname{Zheng}}
\affiliation{QuTech and Kavli Institute of Nanoscience, Delft University of Technology, 2628 CJ Delft, The Netherlands}

\author{Jurgen \surname{Dijkema}}
\affiliation{QuTech and Kavli Institute of Nanoscience, Delft University of Technology, 2628 CJ Delft, The Netherlands}

\author{Nodar \surname{Samkharadze}}
\affiliation{QuTech and Netherlands Organization for Applied Scientific Research (TNO), 2628 CJ Delft, The Netherlands}

\author{Amir \surname{Sammak}}
\affiliation{QuTech and Netherlands Organization for Applied Scientific Research (TNO), 2628 CJ Delft, The Netherlands}

\author{Giordano \surname{Scappucci}}
\affiliation{QuTech and Kavli Institute of Nanoscience, Delft University of Technology, 2628 CJ Delft, The Netherlands}

\author{Lieven~M.~K. \surname{Vandersypen}}
\email[Correspondence to: ]{L.M.K.Vandersypen@tudelft.nl}
\affiliation{QuTech and Kavli Institute of Nanoscience, Delft University of Technology, 2628 CJ Delft, The Netherlands}

\date{August 29, 2020}

\begin{abstract}
Circuit quantum electrodynamics (QED) employs superconducting microwave resonators as quantum buses. In circuit QED with semiconductor quantum-dot-based qubits, increasing the resonator impedance is desirable as it enhances the coupling to the typically small charge dipole moment of these qubits. However, the gate electrodes necessary to form quantum dots in the vicinity of a resonator inadvertently lead to a parasitic port through which microwave photons can leak, thereby reducing the quality factor of the resonator. This is particularly the case for high-impedance resonators, as the ratio of their total capacitance over the parasitic port capacitance is smaller, leading to larger microwave leakage than for 50-$\Omega$ resonators. Here, we introduce an implementation of on-chip filters to suppress the microwave leakage. The filters comprise a high-kinetic-inductance nanowire inductor and a thin-film capacitor. The filter has a small footprint and can be placed close to the resonator, confining microwaves to a small area of the chip. The inductance and capacitance of the filter elements can be varied over a wider range of values than their typical spiral inductor and interdigitated capacitor counterparts. We demonstrate that the total linewidth of a 6.4 GHz and approximately 3-k$\Omega$ resonator can be improved down to 540 kHz using these filters.
\end{abstract}

\maketitle


\section{Introduction} 

Superconducting microwave resonators enable a rich variety of quantum-mechanical phenomena in micro- and nanodevices at cryogenic temperatures, known as circuit quantum electrodynamics (QED). Resonators are used as coupling elements between various types of coherent quantum systems, like superconducting qubits \cite{blais2004,wallraff2004}, electromechanical systems \cite{oaconnell2010,palomaki2013}, collective spin excitations \cite{tabuchi2015,lachance-quirion2020}, and semiconductor quantum-dot (QD) qubits \cite{mi2017, mi2018b,samkharadze2018,landig2018, borjans2020, woerkom2018,landig2019b, scarlino2019a, cubaynes2019a}. 
QD systems typically have a small charge dipole moment, while the coupling to spin qubits is achieved through spin-charge hybridization \cite{cottet2010,hu2012a}. This results in a relatively weak coupling to the resonator mode. High-impedance resonators are therefore desirable, since their small capacitance produces large electric fields that enhance this coupling \cite{samkharadze2016,samkharadze2018,scarlino2019,landig2018}. The same physical advantages of high-impedance resonators have also recently enabled the gate-based rapid single-shot readout of spin states in double quantum dots \cite{zheng2019a}. 

The gate electrodes necessary to form quantum dots in the vicinity of a resonator inadvertently lead to a parasitic capacitance through which microwave photons can leak, thereby reducing the quality factor of the resonator significantly \cite{mi2017a}. This effect is more pronounced for high-impedance resonators, i.e., with impedance in the kiloohm range, as the ratio of their total capacitance over the parasitic capacitance is smaller, leading to larger microwave leakage than for $50 \hyphenohm$ resonators. To mitigate this leakage, symmetric \cite{zhang2014} and dipolar \cite{samkharadze2016} mode resonators have been developed that reduce the mode coupling to the gates, while gate filters \cite{mi2017a} have been employed for popular half- and quarter-wave coplanar resonators with monopolar modes. Until now, the efficiency of gate filters has not been demonstrated in combination with high-impedance resonators that require heavy filtering. Furthermore, current designs have a problematic footprint, including a large interdigitated capacitor and a spiral inductor looping around a bondpad. 

In this work, we develop on-chip filters, consisting of a high-kinetic-inductance nanowire serving as a compact inductor and a small thin-film capacitor, to mitigate leakage from a high-impedance half-wave resonator with silicon double quantum dots (DQDs) at each end. The resonator and inductors are patterned from the same high-kinetic-inductance \ce{Nb_{y}Ti_{1-y}N} film on a \ce{^28Si}/SiGe heterostructure. We use prototyping chips to mimic the parasitic losses by the QD gates with faster fabrication and measurement turnaround than full devices, while minimizing the other lossy mechanisms like dielectric and resistive losses. Finally, we compare thin-film capacitor filters with interdigitated capacitor filters on the aspects of performance, footprint, and integrability.

\section{Methods}

\begin{figure}[tbp]
   \centering
   \includegraphics{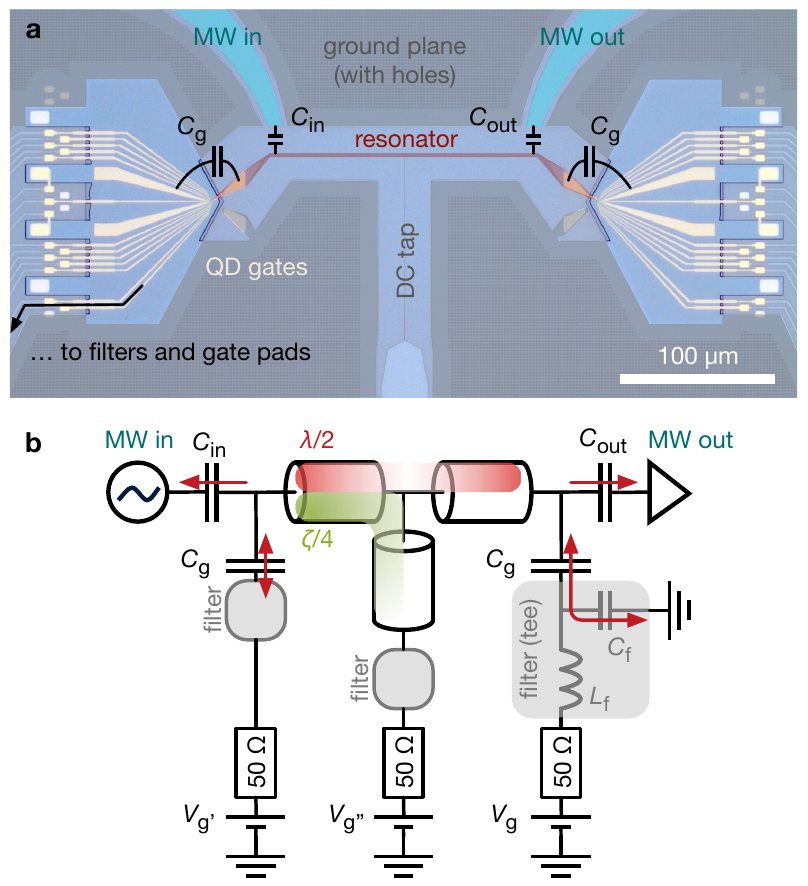} 
   \caption{\figletter{a} Optical image with false-color shading of the central area of the device, showing the superconducting high-impedance nanowire resonator (in red) and the QD gates of a full device (in yellow).     \figletter{b} Simplified electrical circuit of the resonator and its surrounding components. The resonance modes of this device can be understood using three sections of coplanar waveguides with appropriate capacitance and inductance per unit length, $\tilde C_\text{r}$ and $\tilde L_\text{r}$, respectively. A half-wave mode $\lambda/2$ couples the DQDs at each end of the resonator in antiphase, whereas a quarter-wave mode $\zeta/4$ also exists where both DQDs are coupled in phase (only one side of the mode is shown for simplicity). The resonator is probed in transmission through the ``in'' and ``out'' ports with coupling capacitances $C_\text{in}$ and $C_\text{out}$. Because of the physical footprint of the DQD gates at each end, an extra capacitance $C_\text{g} \gg \{C_\text{in}, C_\text{out}\}$ causes the microwave energy to escape from the resonator primarily through the gate fanout lines. To prevent irreversible loss, modeled here by $50 \hyphenohm$ resistors, low-loss microwave filters are fabricated on chip to reflect the microwaves back into the resonator. Filters act as AC grounds through a bias-tee effect. The path of energy escaping the $\lambda/2$ mode is represented by red arrows, with double-ended arrows representing a reflection back into the resonator.}
   \label{fig:device}
\end{figure}

\begin{figure*}[tbp]
   \centering
   \includegraphics{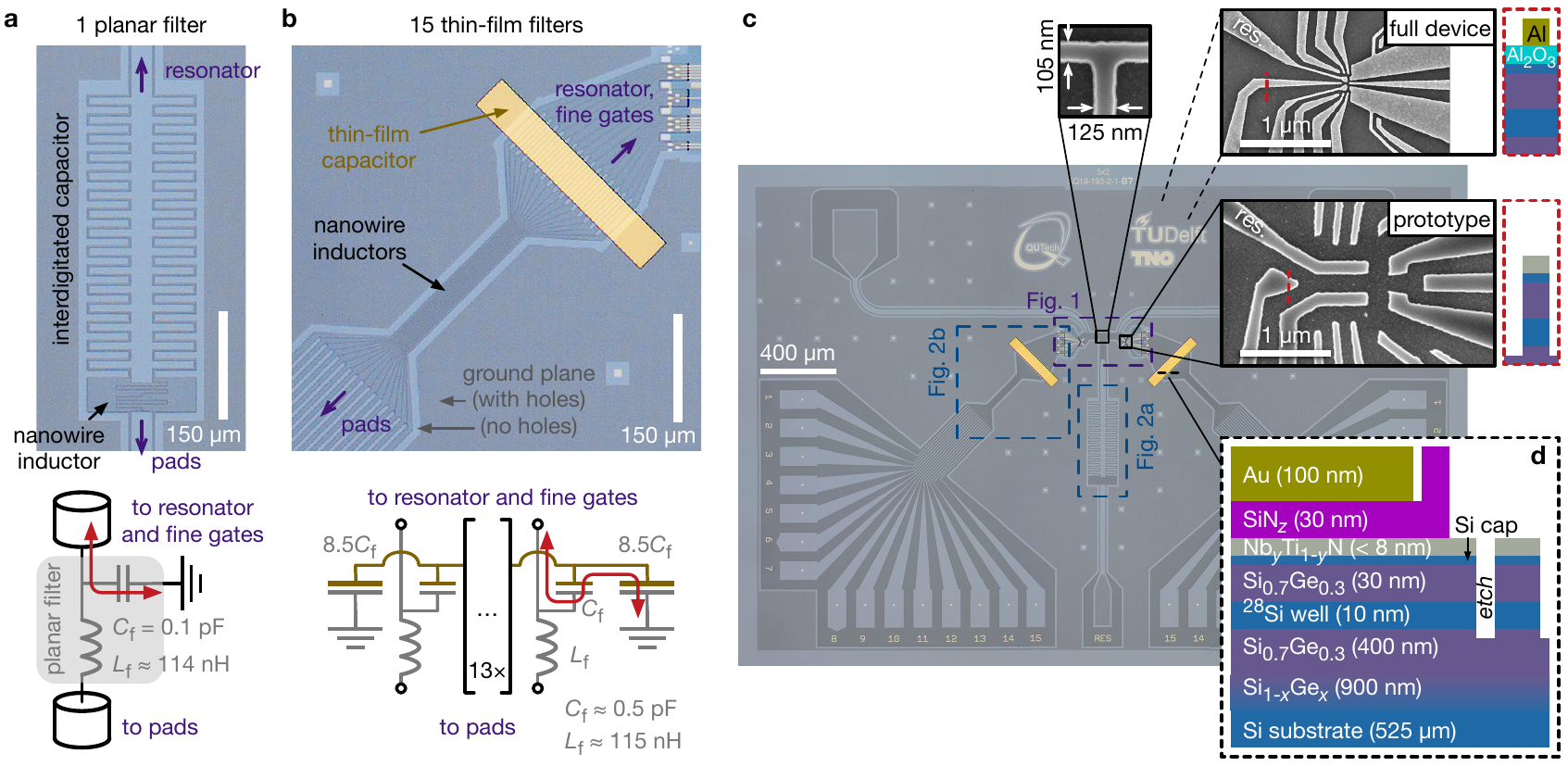} 
   \caption{\figletter{a} Optical image of a planar filter with one line. Each filter can be thought of as an LC microwave bias tee. The inductor $L_\text{f}$ consists of a superconducting high-kinetic-inductance nanowire made from the same film as the resonator. The capacitor $C_\text{f}$ has an interdigitated geometry. Both components are low loss, thanks to superconducting metals and the absence of amorphous dielectrics. The nanowire inductor is much smaller than an equivalent spiral inductor and does not require looping around a bondpad, allowing the filter to be placed closer to the active area. The interdigitated capacitor is still relatively large.     \figletter{b} Optical image of a thin-film filter with 15 gate lines (with the same scale as \figref{fig:filterdesign}a). The thin-film capacitor can be made a lot smaller than its interdigitated equivalent and straddles multiple gate lines at once, thereby dramatically reducing the footprint and simplifying the microwave hygiene. In this implementation, the top capacitor plate is electrically floating to further simplify the integration. The large $8.5C_\text{f}$ series capacitance of the capacitor plate to the ground plane acts as a short for the relevant frequencies.
   \figletter{c} Stitched optical image of a full device chip (4~mm by 2.8~mm) with thin-film filters. To single out the effects from the capacitive loading of the resonator by the gates while minimizing any other loss mechanisms, we use prototyping chips built from the same processed wafers as the full devices, but the quantum-dot areas do not include any of the implanted regions, the gate oxide, or the implant contact pads. The insets show electron microscope images of the nanowire resonator DC tap intersection, the prototyping fine gates mockup and full device fine gates. The mockup and full devices have identical capacitive load, $1.8 \fF$ per DQD.     \figletter{d} Material stack of a thin-film capacitor prototyping chip. See \SupRefSecFab{} for details.}
   \label{fig:filterdesign}
\end{figure*}

The full device being optimized in this work is shown in \figref{fig:device}a. A high-impedance superconducting resonator is etched from a thin, high-kinetic-inductance \ce{Nb_{y}Ti_{1-y}N} film. At each end of the resonator with angular frequency $\w_\text{r}$ and impedance $Z_\text{r}$, a DQD gate structure is fabricated with one accumulation gate attached to the resonator's end, similarly to the device of \refref{samkharadze2018}. The charge displacement in the DQD is then linked to the zero-point root-mean-square voltage swing $V_\text{rms} \propto \w_\text{r} \sqrt{Z_\text{r}}$ at the resonator end through the gate lever arm $\alpha$, allowing a dot-resonator interaction of strength $g_\text{c} \propto \alpha V_\text{rms}$ \cite{beaudoin2016a}. Maximizing this interaction can help reach the strong spin-photon coupling regime \cite{samkharadze2018, mi2018b,landig2018} and increases the sensitivity of the resonator for readout \cite{zheng2019a}. 

This work focusses on reducing the linewidth $\kappa/2\pi$, or improving the quality factor $Q = \w_\text{r}/\kappa$, of the high-impedance resonator using on-chip filters. We model the behavior of this device using the electrical circuit shown in \figref{fig:device}b. The resonator can be roughly approximated as an interrupted coplanar waveguide \cite{goppl2008}, with a half-wave mode $\lambda/2$ and a quarter-wave mode $\zeta/4$. The quarter-wave mode arises from the ``\textsf{T}''-shaped section of the resonator direct current (DC) biasing line terminated by an alternating current (AC) ground provided by the filter (\figref{fig:device}b). With a frequency roughly half of the $\lambda/2$ mode, it is used as a diagnostics tool for the work that follows. The inductance per unit length $\tilde L_\text{r}$ is dominated by the kinetic inductance contribution of the \ce{Nb_{y}Ti_{1-y}N} film section near the current antinode, with nominal sheet inductance $115 \pHpsq$. The nanowire width, in the range 100 to $200 \nm$, serves to adjust the frequency \cite{samkharadze2016}. The kinetic inductance is almost 1000 times larger than the geometric inductance. The effective capacitance per unit length $\tilde C_\text{r}$ is influenced to a large extent by the end sections of the resonator near the voltage antinodes. A typical frequency is $\w_{\lambda/2}/2\pi = 6.4 \GHz$ and $Z_\text{r} \sim 3 \kohm$. The end-to-end length is $l_\text{r} = 250 \um$, which is much smaller than the approximately $9 \mm$ of a coplanar resonator without kinetic inductance. A numerical circuit model and additional details can be found in \SupRefSecModel{}.

We now illustrate why the losses through the gates are increasingly problematic as the resonator impedance increases. We first note that the coupling losses can be approximated in our regime by \cite{mazin2005}
\ma{
	\kappa_\text{g}
	=
	\frac{2}{\pi} \w_\text{r}^3 Z_\text{r} Z_\text{g} C_\text{g}^2 ,
	\label{eq:couplingq}
}
which shows that the losses through a gate fanout $\kappa_\text{g}$, with fixed fanout impedance $Z_\text{g}$, scale as $Z_\text{r}$. For a $3 \hyphenkohm$ resonator, the coupling loss is about 60 times worse than for an equivalent $50 \hyphenohm$ resonator, or 10 times worse than for a $300 \hyphenohm$ one. We use this ideal waveguide formula to get insights into the scaling of leakage but do not rely on quantitative predictions since the gate fanout lines are not simple waveguides. The resonator is probed in transmission, with input and output capacitance $C_\text{in} = C_\text{out} = 0.28 \fF$. The capacitance between each resonator end and the DQD gate ensemble is found to be $C_\text{g} = 1.8 \fF$ using numerical simulations with \textsc{comsol}. Using an equivalent lumped-element parallel LCR oscillator \cite{goppl2008}, we estimate a resonator capacitance of $\tilde C_\text{r} l_\text{r}/2 \approx 8 \fF$ before gate loading. Hence, the gates contribute a significant fraction of the total capacitance, which is a direct side effect of the large bare impedance $Z'_\text{r} = ( \tilde L_\text{r} / \tilde C_\text{r} )^{1/2} \approx 4.0 \kohm$ at fixed $\w_\text{r}$. Given the large contribution of the gates, improved mitigation strategies need to be devised compared with previous work \cite{mi2017a}. The benefit of this large impedance is a large charge-photon coupling strength $g_\text{c}/2\pi \sim 200 \MHz$ \cite{samkharadze2018}. Other work with high-impedance resonators has so far been limited to linewidths $\kappa/2\pi > 10 \MHz$ \cite{scarlino2019,landig2018}, with the exception of \refref{samkharadze2018} where the resonator geometry is not suitable for the coupling of distant qubits. Meanwhile, a reasonable target to achieve two-qubit gates in the dispersive regime would be $\kappa/2\pi < 1 \MHz$ \cite{benito2019c}. This target value is comparable to other order $50 \hyphenohm$ resonators used for spin-photon coupling experiments \cite{mi2017a,cubaynes2019a}, and is also lower than current spin-dephasing rates in current silicon devices with strong coupling \cite{samkharadze2018, borjans2020}.

We propose and demonstrate two models of gate filters to suppress leakage of photons through the gates, which are shown in \figref{fig:filterdesign}. Previous implementations have relied on spiral inductors that loop around bondpads \cite{mi2017a}. Their drawback is that they have a footprint at least as large as a bondpad, and that the inductance values are typically in the tens of nH. The capacitor therefore needs to be large to maintain the LC filter angular cutoff frequency $2\pi f_\text{f} = (L_\text{f}C_\text{f})^{-1/2}$. We advantageously use the high kinetic inductance of the \ce{Nb_{y}Ti_{1-y}N} film to etch low-loss and compact nanowire inductors. Given a target sheet inductance of $115 \pHpsq$, and a nanowire of length $380 \um$ and width $380 \nm$, an inductance of $115 \nH$ can easily be achieved. The resulting planar filter with $f_\text{f} \approx 1.5 \GHz$ is shown in \figref{fig:filterdesign}a. Still, the footprint of typical interdigitated capacitors remains problematic due to their large size, and since extra space has to be allocated between bondpads to allow the ground plane access in between each line. Our target is 15 gate lines per DQD. We therefore also test a thin-film capacitor with \ce{SiN_z} dielectric that straddles 15 gate lines at once, and contacts with the ground plane through a larger series capacitor that acts as a short at the frequencies of interest, as shown in \figref{fig:filterdesign}b. Floating the capacitor top plate is not necessary, but it allows for a single-step liftoff of both the \ce{SiN_z} dielectric and the {Au} metal top plate. As a result, capacitors in the $0.1$ to $1 \pF$ range can be produced with small footprints. Combining the nanowire inductors and the thin-film capacitor, the entire set of filters for 15 lines can fit in the footprint required for a single planar filter. This design also allows us to limit the microwaves to an area much closer to the resonator, simplifying its integration.

In order to test the efficacy of the filters, we use a prototyping chip, which is a simplified version of the full device, as shown in \figref{fig:filterdesign}c. This prototyping chip is made from the same wafers that are used for full devices. All the linewidths reported in this work come from these prototypes. The $100 \hyphenmm$ {\ce{^28Si}/SiGe} wafers are processed with the ion-implanted regions, the 5-to-7-nm-thin \ce{Nb_{y}Ti_{1-y}N} film, the \ce{Al2O3} gate dielectric, and the {Ti}/{Pt} contacts to the implanted regions and the \ce{Nb_{y}Ti_{1-y}N} film; then diced in $20 \hyphenmm$ coupons. Each coupon is further processed with one electron-beam lithography and {\ce{SF6}/He} reactive ion-etching step to define the superconducting elements; optionally one electron-beam lithography and liftoff step to pattern the \ce{SiN_z}/{Au} thin-film capacitor stack (\figref{fig:filterdesign}d); followed by dicing into $4 \mm$ by $2.8 \mm$ individual devices. The pattern is offset such that there is none of the \ce{Al2O3} gate dielectric or the {Ti}/{Pt} contacts used in full devices while using the same starting pieces. Further fabrication details can be found in \SupRefSecFab{}. To accurately capture the effects of the resonator capacitive loading by the gate structure, a simplified version of the gates is patterned directly into the superconducting film, as shown in \figref{fig:filterdesign}c. This structure has the same capacitance to the resonator as the real gates, according to numerical simulations with \textsc{comsol}. Hence, the resonator losses should be dominated by microwave leakage into the gate fanout, as opposed to dielectric or resistive losses. The devices are then measured in a \ce{^3He} refrigerator with a base temperature of approximately $270 \mK$, unless otherwise specified. Setup details can be found in \SupRefSecSetup{}. The gate pads of each DQD are wirebonded to each other, and then to a common port with a $50 \hyphenohm$ termination (one line per DQD side) on a five-port printed circuit board (PCB). This simulates the irreversible loss of microwaves in a dilution refrigerator with resistive-capacitive filters and instruments attached to each gate line. Linewidth analysis details can be found in \SupRefSecAnalysis{}.

\section{Results}

\subsection{Planar filters}

\begin{figure*}[tbp]
   \centering
   \includegraphics{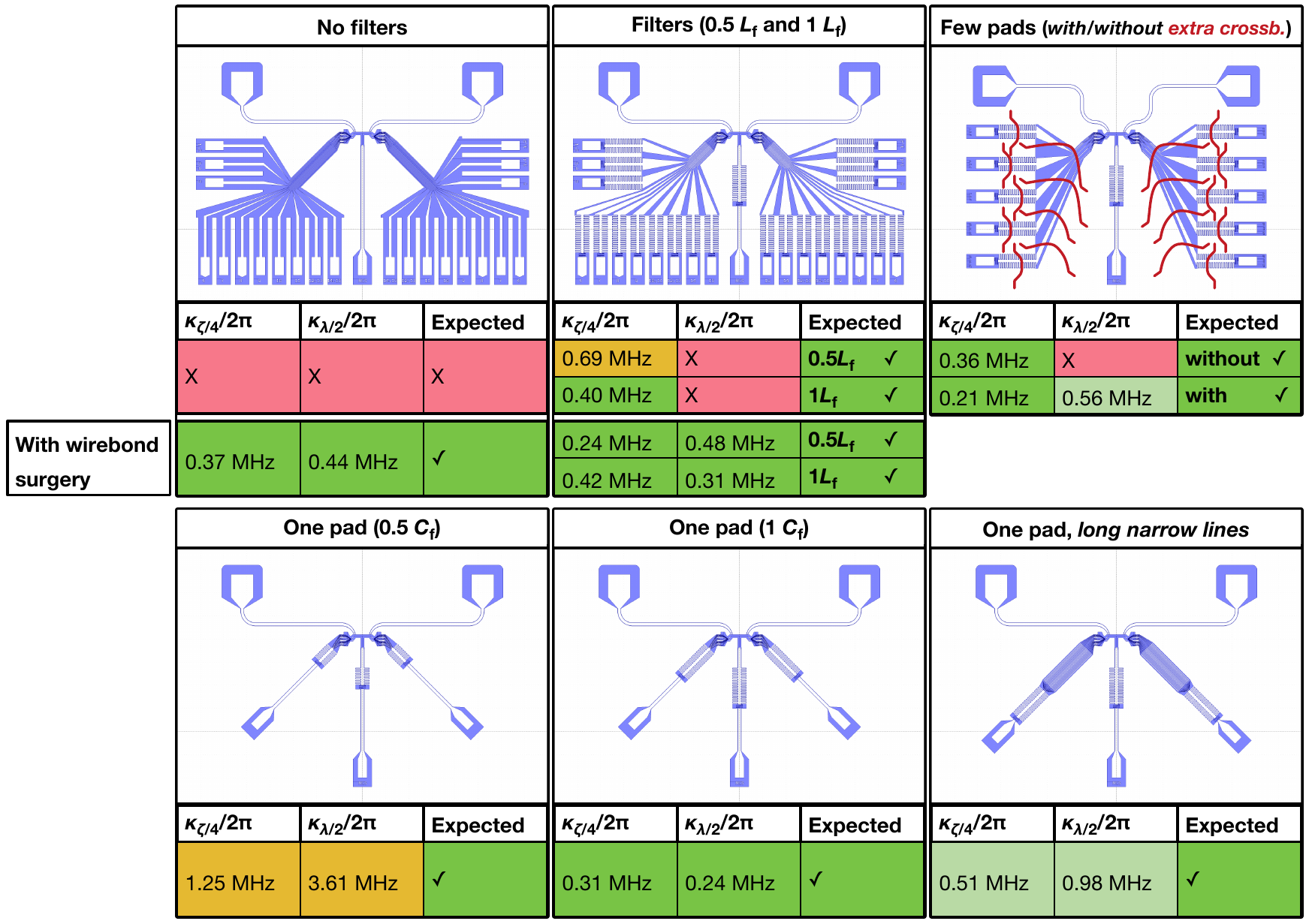} 
   \caption{Summary for planar filter prototypes. The $\zeta/4$ mode frequencies lie between 3 and $4 \GHz$, while the $\lambda/2$ mode frequencies lie between 6 and $7.5 \GHz$. The experimental splits are designed to separate the problems caused by insufficient or defective filtering from those caused by poor microwave hygiene. For each prototype, up to three variants are tested. A ``no gates'' variant, not shown in the table, is meant as a control experiment to measure frequencies and linewidths in the absence of gate loading. Typical values yield $\kappa_{\zeta/4}/2\pi = 0.3 \MHz$ and $\kappa_{\lambda/2}/2\pi = 0.9 \MHz$ for the two modes. The ``with gates'' variant mimics the full devices. In certain cases, after initial measurements, extra wirebonds shorting the gates to the ground plane are added close to the resonator area and chips are then remeasured, resulting in the ``with wirebond surgery'' variant. This sanity check procedure is useful to verify that the failure of chips is due to insufficient filtering or poor microwave hygiene, and not due to other problems like a resonator defect. The color coding is a subjective assessment of whether or not the linewidth is optimal, with green being very close to ideal ($\lesssim 0.5 \MHz$), yellow being not ideal but still $\lesssim 4 \MHz$, and red being $> 10 \MHz$. For this set, $L_\text{f} \approx 114 \nH$ and $C_\text{f} \approx 0.1 \pF$. The `expected' column is a binary assessment of whether the linewidth should be narrow (\checkmark) or broad (\textsf{X}) based on the presence or absence of filtering. See the main text for discussion.}
   \label{fig:tableplanar}
\end{figure*}

We now turn to the results for devices with planar filters, which are summarized in \figref{fig:tableplanar}. We first validate the experimental protocol with various consistency checks. We verify that the devices measured in our \ce{^3He} system have similar linewidths, both the good and poor ones, to the ones obtained in our dilution refrigerator setup with all gate lines connected individually to real instruments. Second, we measure ``no filters'' devices and find that the linewidth is so broad that the resonances can barely be found, and at times cannot be seen at all. This usually means that the linewidth is $\gtrsim 15 \MHz$. This is to be compared with the coupling linewidths $\kappa_{\zeta/4}^\text{ext}/2\pi \sim 0.03 \MHz$ and $\kappa_{\lambda/2}^\text{ext}/2\pi \sim 0.2 \MHz$, estimated from numerical simulations. As the resonator is usually undercoupled, the improvements in linewidths are also visible in the larger transmission amplitude. A ``no gates'' variant, not shown in the figure, is meant as a control experiment to measure frequencies and linewidths in the absence of gate loading. Typical values yield $\kappa_{\zeta/4}/2\pi = 0.3 \MHz$ and $\kappa_{\lambda/2}/2\pi = 0.9 \MHz$ for the two modes. Because of the device-to-device variability in resonance frequency, it is difficult to precisely measure the gate loading (the difference in frequency between the ``no gates'' and ``with gates'' prototypes). The variability is due partly to the thicker superconducting film in the wafer center, and partly to the lithographic variability of long narrow features. Nevertheless, we usually see a $0.5$- to $1 \hyphenGHz$ frequency difference between the ``no gates'' and ``with gates'' prototypes, in line with our estimates from numerical simulations. Finally, a very useful consistency check is the ``wirebond surgery'' technique. This consists of adding extra wirebonds to a previously measured device to diagnose the cause of the failure or suboptimal linewidth. It is usually possible to short the gate lines directly to the ground plane before the filters, and even a few hundreds of microns from the resonator, to effectively remove the gate fanout losses. This useful technique allows us to confidently identify failure mechanisms due to filters, as opposed to an accidental failure of the resonator for example, with minimal work.

Next, we look at the various prototypes shown in \figref{fig:tableplanar}. The experimental splits are designed to separate the problems caused by insufficient or defective filtering from those caused by poor microwave hygiene. Because of the large kinetic inductance of the superconducting film, certain waveguides or parts of the gate fanout lines can have associated wavelengths that are problematic at the frequencies of interest, effectively causing spurious resonances at a scale that would be otherwise unexpected. Another potential problem can be the finite inductance between different ground-plane sections causing out-of-phase return currents that hinder the functioning of components. These problems are generically referred to as microwave hygiene problems. To keep the fabrication process simple and maintain magnetic field compatibility, we opt not to locally deposit a thicker ground plane, and to avoid the use of air bridges. The different ground-plane sections are always connected with crossbonds, as is common practice. In the case of the ``few pads'' prototypes, we specifically test extra crossbonds between the capacitors.

The most striking feature seen in the top row of \figref{fig:tableplanar} is that the filters seem to be somewhat effective for the quarter-wave mode, but not for the half-wave mode. We attribute this to a microwave hygiene problem, where the filters are only effective for the quarter-wave mode because of its lower frequency. The ``one pad'' prototypes in the second row mean to test the design of a single planar filter while ensuring proper microwave hygiene. Therefore, all gates are attached to the same filter unit and surrounded by a well connected ground plane. Here, we find that the prototype with $0.5 C_\text{f}$ has larger linewidths than the prototype with $1 C_\text{f}$, which we attribute to a difference in filtering efficacy. To further our understanding, we also test a variant that has longer sections of narrow gate lines before the filter (``one pad, long narrow lines''). We find that the linewidths are not as small as our best performing ``one pad'' device, but still considered acceptable. We think it is possibly due to the distance between the resonator and filter that interferes with the filtering. Notably, the linewidths of good prototypes with gates are consistently smaller than those without gates, an effect that we attribute to the larger total capacitance, hence lower impedance and frequency, of gated prototypes. 

\begin{figure*}[tbp]
   \centering
   \includegraphics{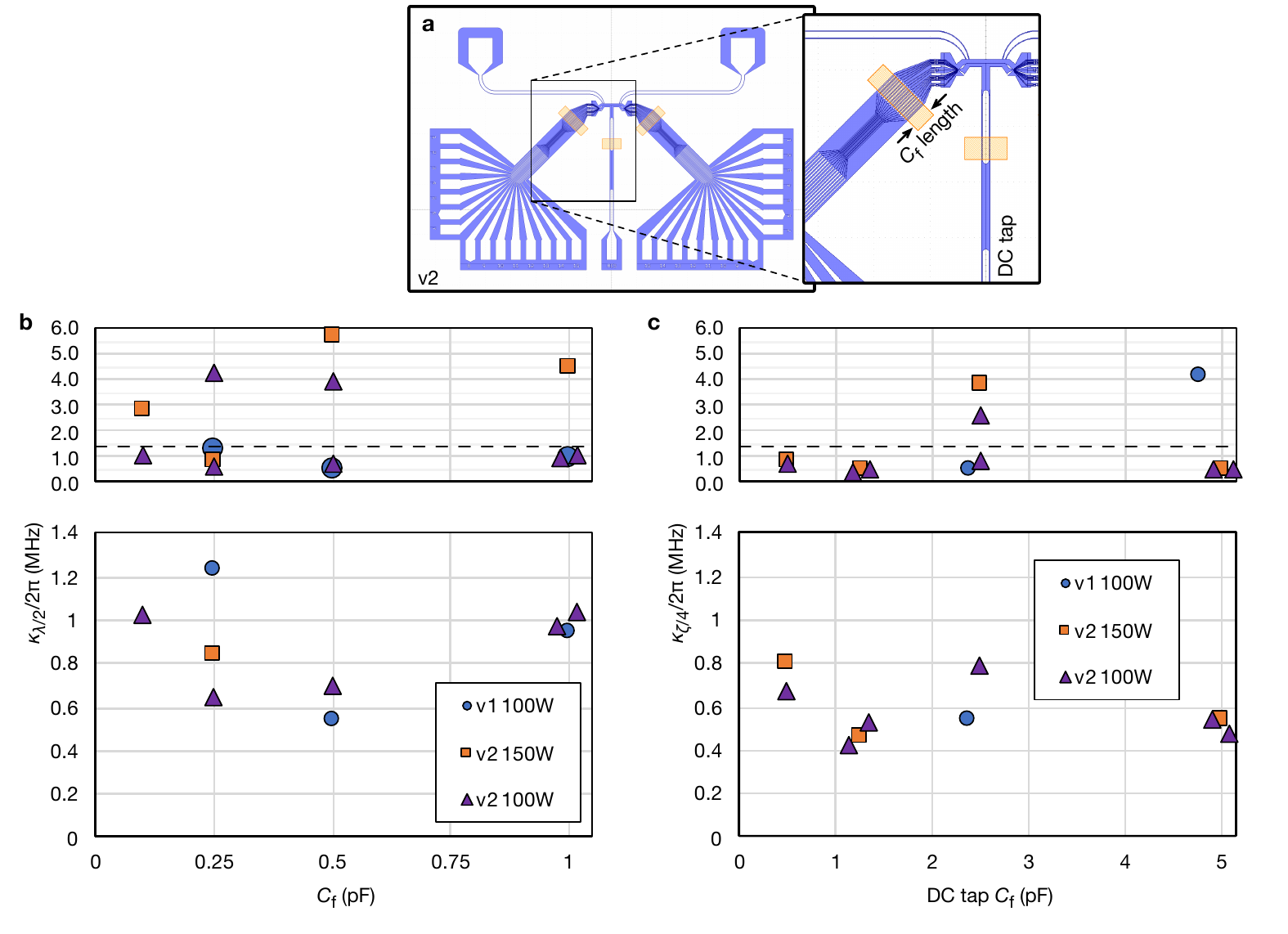} 
   \caption{Summary for thin-film filter prototypes.     \figletter{a} The experimental splits are designed to test the impact of the capacitor area, by changing the length of the capacitor (as shown in the inset), and include results from two very slightly different designs (v1 and v2) and \ce{SiN_z} deposition powers ($100 \watt$ and $150 \watt$). The capacitor $C_\text{f} = 1 \pF$ has a width of $6 \um$ (top plate width of $320 \um$) and length of $100 \um$, while the \ce{SiN_z} thickness is $30 \nm$.    \figletter{b} The $\lambda/2$ mode frequencies lie between 6 and $7.5 \GHz$. Two groups are observed: good devices with $\kappa_{\lambda/2}/2\pi < 1.25 \MHz$, and poor devices with $\kappa_{\lambda/2}/2\pi > 2 \MHz$. The reason for their failure is unknown, but in two cases we observe that the resonance is recovered using wirebond surgery with linewidths $1.1 \MHz$ and $0.7 \MHz$. The best performing device achieved $\kappa_{\lambda/2}/2\pi = 0.540 \MHz$, or a quality factor of $11\,900$.     \figletter{c} Linewidths for the $\zeta/4$ mode. The DC tap capacitor for this mode is scaled up in size compared with the gate line capacitors to improve the AC grounding of the mode (see main text). While the half-wave mode is insensitive to losses through the DC tap because of its symmetry, the quarter-wave mode can lose energy through both the gate lines and the DC tap, making the contributions from the different filters convoluted for this mode. In all plots, some points are slightly offset horizontally for clarity.}
   \label{fig:thinfilm}
\end{figure*}

The first row ``few pads'' prototype means to test the microwave hygiene further in the case where a bigger chip size would ultimately be adopted. The hypothesis is that the failure of the ``filters'' prototypes from the first row comes from the insufficient space between the gate pads. Because of the high kinetic inductance, the sections connecting the interdigitated capacitors to the rest of the ground plane act as inductors, hindering their action. Results show that the extra space allowed between the pads in the ``few pads'' prototype does not help. However, adding crossbonds to further distribute the ground-plane potential improves the linewidth of the high-frequency mode to a good level. This seems an acceptable design with efficient filters, with the caveat that the footprint allowed a maximum of 5 or 6 gate lines per DQD given our chip size. 

Finally, we note that the results for the quarter-wave mode linewidths in the $0.5 L_\text{f}$ and $1 L_\text{f}$ variants are in qualitative agreement with those of the $0.5 C_\text{f}$ and $1 C_\text{f}$ variants, but they are convoluted with the microwave hygiene issue aforementioned.

\subsection{Thin-film filters}

In order to find a more extensible solution to the filtering problem, we turn to a thin-film capacitor design, shown in \figref{fig:thinfilm}a. This design reaches the target of 15 gate lines per DQD. An example device is shown in \figref{fig:filterdesign}c, and the filter operation is described in \figref{fig:filterdesign}b. For experimental splits, we change the area of the $C_\text{f}$ capacitor, as well as the deposition conditions for the \ce{SiN_z} film. We also cool down several instances of each prototype. One reason for this extensive testing is to verify whether there is a critical area beyond which the dielectric losses in the thin-film capacitor would degrade the quality of the filtering. From a simple lumped-element circuit model perspective, the larger the capacitance is, the more efficient the filtering should be. In practice however, it is useful to only use the minimum amount of filtering (i.e., highest $f_\text{f}$ possible), since this allows control signals to the DQDs with less distortion.
We find that the size of the capacitor does not have a large effect on the linewidth. Most devices perform acceptably with linewidths $< 1.25 \MHz$. However, a few devices have linewidths that are significantly broader than this. In two of those devices, wirebond surgery successfully recovered a linewidth comparable to the best devices. This seems to indicate a problem with the filters. This type of failure is not observed in later fabrication rounds, but we include the results here for completeness. The detailed results are shown in \figref{fig:thinfilm}b. There seems to be an optimal point at $C_\text{f} = 0.5 \pF$ where the best devices have the narrowest linewidths, but considering the spread of the results, we cannot be certain that this is a systematic effect. 

The $\zeta/4$ mode performs similarly to the $\lambda/2$ mode with approximately $30\pc$ narrower linewidths, as shown in \figref{fig:thinfilm}c. For the v2 design, shown in \figref{fig:thinfilm}a, the capacitor on the DC tap is larger by a factor 5 than the gate line ones. This is done to improve the AC grounding of the mode. The ratio is 9.5 for the v1 design (not shown). While the half-wave mode is insensitive to losses through the DC tap because of its symmetry, the quarter-wave mode can lose energy through both the gate lines and the DC tap, making the contributions from the different filters convoluted. Therefore, the linewidth results for this mode are provided for completeness, but are not factored into the optimization process. 

Given that the thin-film solution produces devices with $<1 \MHz$ linewidths with the right number of gate lines, we are satisfied with these results. It is worth noting that the cutoff frequency of each line can be adjusted individually, by changing the capacitance or the nanowire inductance, simply by adjusting the widths of the gate-line sections.

\section{Conclusion}

In summary, we demonstrate compact on-chip filters for high-impedance resonators that prevent the losses of microwave energy through the gate lines of the coupled QD structure. The inductors are made of the same high-kinetic-inductance superconductor as the resonator. This produces small inductors of large inductance that can be placed anywhere on the chip, as opposed to spiral inductors. We compare two approaches to implement the filter capacitor: one with a planar interdigitated capacitor and one with an overlapping thin-film capacitor. The planar filters performed well when used with sufficient crossbonds; however, their footprint is relatively large, making the solution inconvenient as the number of gate lines increases. The thin-film capacitors are fabricated with a single additional lithography step and dramatically reduce the total footprint of the filter. Our implementation has one capacitor plate overlapping 15 gate lines, effectively producing a very compact filter unit. When combined with the nanowire inductors, this simplifies the microwave engineering by confining the resonator energy to a small area of the chip. We demonstrate that the total linewidth of a $6.4 \hyphenGHz$ resonator can be improved down to $540 \kHz$ using these filters, therefore achieving a loaded quality factor of $11\,900$. It is understood that the best solution depends on the combination of footprint and linewidth requirements. For us, the thin-film solution is the only one to satisfy both. With these filters in place, the biggest source of loss in full devices is then dominated by the gate resistance and dielectric losses of the QD area, which will be addressed in future work. Since the resonator and its ground plane have been shown to be compatible with in-plane magnetic fields up to $6 \tesla$ \cite{samkharadze2016}, we do not expect a different behavior in the current case. These low-loss resonators with large coupling to quantum dots could allow more sensitive hybrid spin-superconducting devices to realize long-range two-qubit gates, high-speed gate-based readout, circuit QED experiments with single spins, as well as more fundamental experiments in the device and materials fields.

\begin{acknowledgments}

The authors thank L.\ DiCarlo for useful discussions, L.\ DiCarlo and his team for access to the \ce{^3He} cryogenic measurement setup, L.~P.\ Kouwenhoven and his team for access to the \ce{Nb_{y}Ti_{1-y}N} film deposition, F.\ Alanis Carrasco for assistance with sample fabrication, and other members of the spin qubit team at QuTech for useful discussions.
This research was undertaken thanks in part to funding from the European Research Council (ERC Synergy Quantum Computer Lab) and the Netherlands Organization for Scientific Research (NWO/OCW) as part of the Frontiers of Nanoscience (NanoFront) programme.

\end{acknowledgments}

\section*{Author contributions}
\PHC\ and G.Z.\ conceived and planned the experiments.
G.Z.\ and J.D.\ performed the electrical cryogenic measurements.
J.D.\ performed numerical simulations.
\PHC\ designed the devices, and N.S.\ provided advice.
\PHC\ and J.D.\ fabricated the devices.
A.S.\ contributed to sample fabrication.
A.S.\ grew the heterostructure with G.S.’s supervision.
\PHC, G.Z., J.D.\  and L.M.K.V.\ analyzed the results.
\PHC\ wrote the manuscript with input from all co-authors. 
L.M.K.V.\ supervised the project.

\section*{Data availability}
The data reported in this paper are archived online at \url{https://dx.doi.org/10.4121/uuid:913e3aaf-71ac-4a00-b191-0ab8df56280c}.

\appendix

\newcommand{\RefFigDevice}{\figref{fig:device}} 
\newcommand{\RefFigDesign}{\figref{fig:filterdesign}} 
\newcommand{\RefEqCouplingQ}{\eqnref{eq:couplingq}} 

\section{\uppercase{Device fabrication}}
\label{sec:fabrication}

\begin{figure}[tbp]
   \centering
   \includegraphics{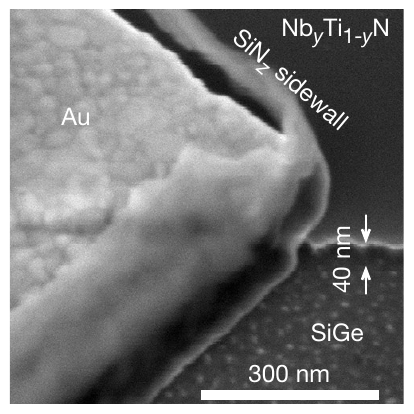} 
   \caption{Angled-view scanning electron microscope image of the thin-film capacitor structure overlapping a step and the \ce{Nb_{y}Ti_{1-y}N} film.}
   \label{fig:tfsem}
\end{figure}

The {\ce{^28Si}/SiGe} quantum-well heterostructure is grown on a $100 \hyphenmm$ {Si} wafer via reduced-pressure chemical vapor deposition, as per \RefFigDesign{}d. 
Photolithography alignment markers are plasma etched into the surface with a {Cl/HBr} chemistry. 
Doped contacts to the quantum well are formed by \ce{^31P} implantation and activated with a $700 \degC$ rapid thermal anneal.
The $5{-}7 \nm$ superconducting {\ce{Nb_{y}Ti_{1-y}N}} film is deposited via magnetron sputtering, preceded by a hydrofluoric acid dip and Marangoni drying, and followed by liftoff of the resist-covered quantum dot areas. The sheet inductance is targeted to be around $115 \pHpsq$. 
The $10 \nm$ \ce{Al2O3} gate oxide is grown by atomic layer deposition, followed by wet etching with buffered hydrofluoric acid everywhere except for the resist-covered quantum-dot areas.
Contacts to implants, contacts to the \ce{Nb_{y}Ti_{1-y}N} film, and electron-beam-lithography alignment markers are patterned with {Ti}/{Pt} evaporation preceded with buffered hydrofluoric acid dip and followed by liftoff.
The wafer is diced into pieces for further electron-beam-lithography steps.
The \ce{Nb_{y}Ti_{1-y}N} film is etched via \ce{SF6}/{He} reactive ion etching to define the resonator, inductors, capacitors, and gate lines in a single electron beam lithography step, leaving a $40 \hyphennm$ step after the etch.
The thin-film capacitor is patterned by first sputtering $30 \nm$ of silicon nitride in a conformal deposition, then evaporating $5 \nm$ of {Ti} and $100 \nm$ of {Au} in a directional deposition, allowing for a single patterning and liftoff step. The \ce{SiN_z} conformal deposition covers the $40 \nm$ steps created during the etch of the \ce{Nb_{y}Ti_{1-y}N} film. The resulting structure is shown in \figref{fig:tfsem}. The \ce{SiN_z} relative dielectric constant is not measured, and is estimated to be approximately $6$ based on typical values for sputtered \ce{SiN_z}. The top-plate metal is chosen sufficiently thick to cover the steps and have low electrical resistance.
Pieces are diced into individual device chips for electrical characterization.

\section{\uppercase{Numerical resonator model}}
\label{sec:model}

In this section, we present a numerical method to model the resonator's half-wave and quarter-wave modes, together with the effect of the filters, as shown in \RefFigDevice{}. The model can easily be adapted with different levels of complexity to better capture the effects of the various impedances of the different waveguide and resonator sections, while remaining computationally fast by avoiding three-dimensional (3D) microwave simulations. Simulations with Sonnet are also performed on individual components (like the resonator or a planar filter) as a consistency check, but the results are not presented in this work. As seen from the optical image in \RefFigDevice{}a, the resonator does not have a simple coplanar waveguide geometry. However, we can get a good (yet still relatively simple) model of it by using combinations of coplanar waveguide sections. These sections then account, to a better degree, for the spatially inhomogeneous capacitance and inductance per unit length of the system.

\begin{figure*}[tbp]
   \centering
   \includegraphics{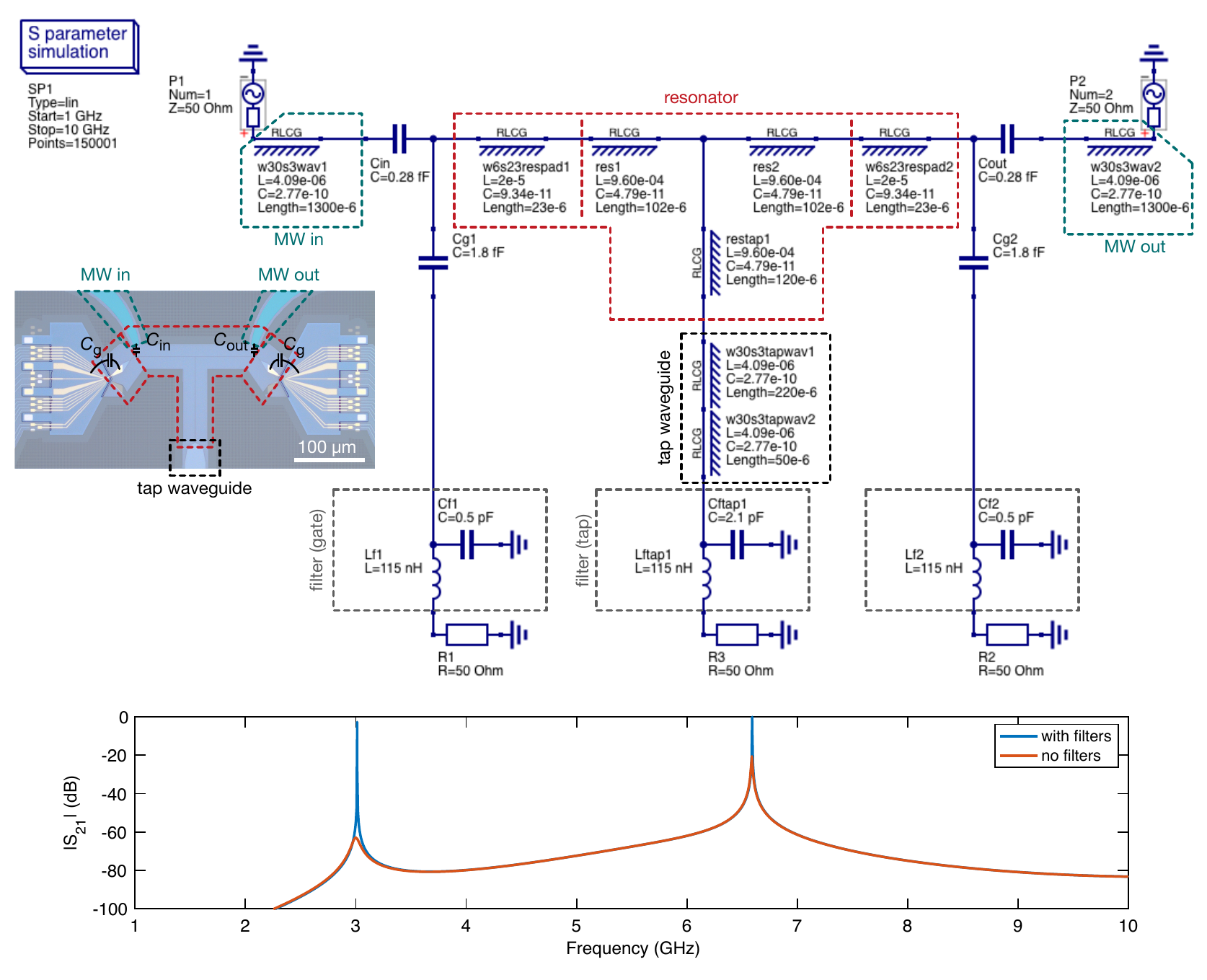} 
   \caption{Numerical model of the resonator and gate filters implemented using the software \textsc{qucs}. RLCG elements represent waveguides with arbitrary inductance per unit length \texttt{L} and capacitance per unit length \texttt{C} (in SI units). The corresponding device elements are delimited by dashed boxes. The model includes effects from the diamond-shaped pads at the end of the resonator narrow section, capacitive loading by the gates, and various waveguide impedances. In some cases, the waveguide dimensions $w$ and $s$ are indicated in the element name in units of microns for convenience. The narrow resonator sections \texttt{res1}, \texttt{res2}, and \texttt{restap1} have $w = 120 \nm$ and $s = 17 \um$. The diamond-shaped pad \texttt{w6s23respad1} has $w = 6 \um$ and $s = 23 \um$. While the bare impedance of the narrow resonator section consisting of \texttt{res1} and \texttt{res2} is quite high at about $4.5 \kohm$, the resonator is so small that the capacitive loading by the surrounding elements brings the effective impedance down to approximately $3.2 \kohm$. This last value is obtained by replacing the \texttt{w6s23respad1}, \texttt{res1}, and \texttt{Cg1} elements, and their symmetric counterparts, with a single RLCG element of the same total length, fixing $\texttt{L} = 9.60\text{e-4}$ and yielding $\texttt{C} = 9.4\text{e-11}$.}
   \label{fig:modelA}
\end{figure*}

The model is implemented using the open source software \textsc{qucs} (\url{https://sourceforge.net/projects/qucs/}, \mbox{v0.0.19}). It makes use of a mixture of lumped components, RLCG waveguide components, and performs a S-parameter simulation. The circuit is shown in \figref{fig:modelA}. To calculate the inductance per unit length $\tilde L$ and capacitance per unit length $\tilde C$ of the different coplanar waveguide sections, we use the analytical formulae
\ma{
	\tilde L
	&=
	\frac{\mu_0}{4}\frac{K(k')}{K(k)} + \frac{L_\text{k}}{w} 
	\label{eq:wavind} , \\
	\tilde C
	&=
	4 \epsilon_0 \epsilon_\text{eff} \frac{K(k)}{K(k')} ,
	\label{eq:wavcapa}
}
where $\mu_0$ is the permeability of free space, $\epsilon_0$ is the permittivity of free space, $\epsilon_\text{eff} = (11.7+1)/2$, $K$ is the complete elliptic integral of the first kind, $k = w/(w+2s)$, $k' = \sqrt{(1-k^2)}$, $L_\text{k}$ is the sheet inductance, and $w$ and $s$ are the center conductor width and gap width, respectively.

In a real-use case, the capacitances in the model are determined either by \textsc{comsol} simulations or by the waveguide geometry \eqnref{eq:wavcapa}. The resonator width is measured with a scanning electron microscope. The only free parameter is then $L_\text{k}$, on which the different values for $\tilde L$ depend. We determine $L_\text{k}$ by adjusting its value so that the two resonance frequencies match the experimental ones. $L_\text{k}$ varies from wafer to wafer, and also from center to edge within one wafer. The model accurately describes the two resonance frequencies simultaneously. However the linewidths show only qualitative agreement with the measured ones. This could be due to unaccounted factors: for example, to dielectric or resistive losses in the resonator and filters, to the overly simplistic description of the ports' impedances [see \RefEqCouplingQ{}], or to other 3D microwave effects. 

The model presented in \figref{fig:modelA} can be further refined with little computational overhead to include many gate channels per DQD, to describe the floating top capacitor plate in the thin-film filter implementation, to change lumped elements into distributed RLCG ones, or to attempt to model the effect of the finite impedance of the gate lines on the filtering efficacy. We try various combinations of these refinements. They sometimes help identify undesirable features, like resonances in gate fanout lines or in other waveguides. The outcomes serve as a quick design starting point. However, we find that chip-scale effects can significantly degrade the predicted performance, as explained in the main text.

\section{\uppercase{Measurement setup}}
\label{sec:setup}

\begin{figure}[tbp]
   \centering
   \includegraphics{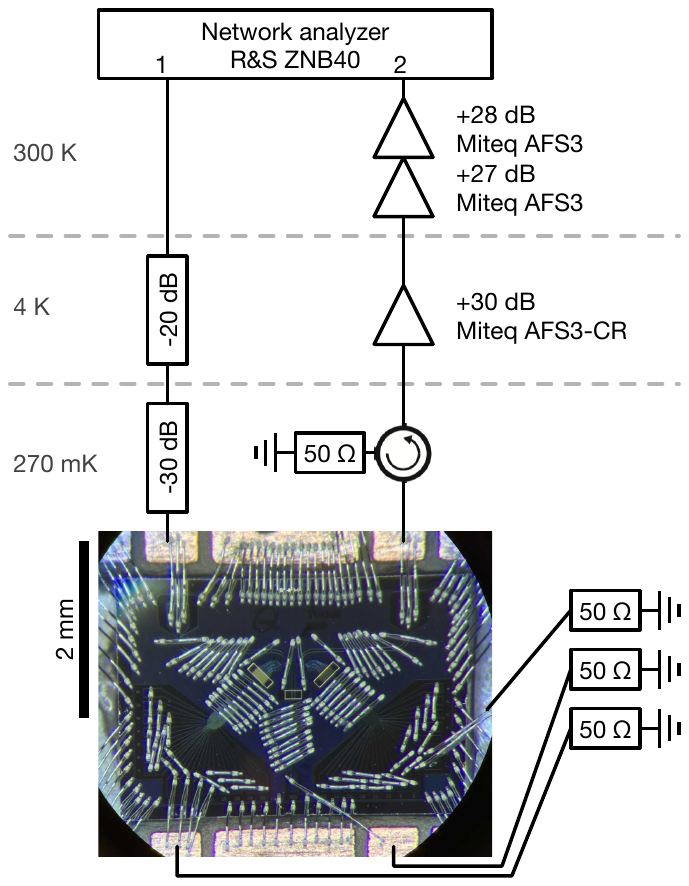} 
   \caption{Measurement setup for the \ce{^3He} system. The picture shows a wirebonded device with crossbonds mounted on the PCB.}
   \label{fig:setup}
\end{figure}

The measurement setup is shown in \figref{fig:setup}. 

\section{\uppercase{Data analysis}}
\label{sec:analysis}

We generally observe a slight power dependence of the linewidths. The narrower linewidths are typically $5 \pc$ to $30 \pc$ broader at $-110 \dBm$ power than at higher powers. Since we are interested in the low photon-number regime, all linewidths are measured with $-110 \dBm$ delivered at the PCB, except for some of the broadest ones where the signal-to-noise ratio is too small. Although a Lorentzian lineshape typically yields acceptable fits for $\kappa/2\pi \lesssim 1 \MHz$, the broader resonances are better captured by a Fano lineshape:
\ma{
	\abs{S_{21}}^2
	&=
	a\abs{\frac{(\w-\w_\text{r})/q + \kappa^\text{ext}/2}{i(\w-\w_\text{r}) + \kappa/2}}^2 ,
}
where $a$ is an arbitrary parameter, $q$ is a complex Fano factor, and $\kappa = \kappa^\text{ext}+\kappa^\text{int}$. The fits do not allow us to independently determine the external and internal losses, $\kappa^\text{ext}$ and $\kappa^\text{int}$, respectively.

\bibliographystyle{apsrev4-1-title} 
\bibliography{/Users/pharveycollard/OneDrive/Papers/PHC-cloud}

\end{document}